\documentclass[letterpaper,twocolumn]{jpsj3}
\usepackage{txfonts}
\usepackage{color}

\title{Gapped Spin Excitation in Magnetic Ordered State \\ on Yb-Based Zigzag Chain Compound YbAgSe$_{2}$}

\author{Fumiya Hori$^1$\thanks{hori.fumiya.36s@st.kyoto-u.ac.jp}, Shunsaku Kitagawa$^1$, Kenji Ishida$^1$, \\ Souichiro Mizutani$^2$, Yudai Ohmagari$^2$ and Takahiro Onimaru$^2$}
\inst{$^1$Depertment of Physics, Kyoto University, Kyoto 606-8502, Japan \\
$^2$Department of Quantum Matter, Graduate School of Advanced Science and Engineering, Hiroshima University,
Higashihiroshima 739-8530, Japan \\
} 

\abst{We report the $^{77}$Se-nuclear magnetic resonance (NMR) results of trivalent Yb zigzag chain compound YbAgSe$_2$, which is a sister compound of YbCuS$_2$.
The $^{77}$Se-NMR spectrum was reproduced by considering two different Se sites with negative Knight shifts and three-axis anisotropy.
Above the Néel temperature $T_{\rm N}$, the Knight shift is proportional to the bulk magnetic susceptibility.
Below $T_{\rm N}$, the extremely broad signal with weak intensity and the relatively sharp signal coexist, suggesting that one is strongly influenced by internal magnetic fields and the other remains relatively unaffected by these fields in the magnetic ordered state.
The nuclear spin-lattice relaxation rate $1/T_1$ remains almost constant above $T_{\rm{N}}$ and abruptly decreases below $T_{\rm{N}}$.  
In contrast to YbCuS$_2$, a $T$-linear behavior of $1/T_1$ at low temperatures was not observed at least down to 1.0~K in YbAgSe$_2$.
Our results indicate that the gapless excitation is unique to YbCuS$_2$, or is immediately suppressed in the magnetic fields.}

\begin{document}
\maketitle

\section{Introduction}

In recent years, there has been significant attention and interest in novel charge-neutral quasiparticles observed in strongly correlated insulators.~\cite{YbMgGaO4-1, YbMgGaO4-2, YbMgGaO4-3, NaYbSe2, NaYbSe2_spinon_fermi_surface, YbB12_resistivity, YbB12, YbB12_2, YbIr3Si7, YbIr3Si7_NMR, Hori2022, Hori2023}
The gapless excitations of such quasipartices, which are seemingly inconsistent with the bulk charge-band gap observed in transport measurements, have been reported in some rare-earth based compounds; Yb-based frustrated insulating magnets YbMgGaO$_4$~\cite{YbMgGaO4-1, YbMgGaO4-2, YbMgGaO4-3} and NaYbSe$_2$~\cite{NaYbSe2, NaYbSe2_spinon_fermi_surface}, Kondo insulators YbB$_{12}$~\cite{YbB12_resistivity, YbB12, YbB12_2} and YbIr$_3$Si$_7$~\cite{YbIr3Si7, YbIr3Si7_NMR}.
 YbCuS$_2$ is one of such rare-earth-based compounds, where the Yb$^{3+}$ ions with magnetic moments form zigzag chains along the $a$ axis~\cite{YbCuS2structure}.
 Due to the competing interactions between nearest and next-nearest neighbors in the zigzag chains, magnetic frustration is expected in YbCuS$_2$~\cite{Majumdar, zigzag1, zigzag2, zigzag3, Saito1, Saito2, Saito3}. 
 At zero magnetic field, the specific heat exhibits a sharp peak at $T_{\rm{N}} \sim 0.95$~K~\cite{Ohmagari1, Ohmagari2}.
In the previous study~\cite{Hori2023}, we performed $^{63/65}$Cu-nuclear quadrupole resonance (NQR) measurements and revealed the first-order magnetic transition at $T_{\rm{N}}$ and an incommensurate magnetic structure with much smaller Yb magnetic moments than the anticipated one from the crystalline electric field ground state.
In addition, we found the gapless excitations in YbCuS$_2$, suggesting the presence of the charge-neutral quasiparticles.
Although the magnetic frustration arising from the Yb zigzag chains seems to play a crucial role in leading to this ordered state and the gapless excitations~\cite{Hori2023, Hori2022}, the origin and the characteristics of these phenomena still remain unclear.
Recently, it has been theoretically proposed that the anisotropic exchange interactions originating from the Yb zigzag chains may lead to gapless excitations~\cite{Saito1, Saito2, Saito3}.
Therefore, the investigation on other Yb-based zigzag-chain compounds is highly desired.

Yb-based compound YbAgSe$_2$ has an orthorhombic structure with the space group $P2_1 2_12_1$ (No.~19,~$D^4_2$) and the Yb zigzag chains are formed as shown in Fig.~\ref{simulation}(a)~\cite{YbAgSe2_structure}, similar to YbCuS$_2$~\cite{YbCuS2structure}. 
There are two crystallographically different Se sites.
The value of the effective magnetic moment estimated from the magnetic susceptibility above 40~K is 4.77~$\mu_{\rm{B}}$, which is close to that expected for the trivalent Yb$^{3+}$ ions (4.54~$\mu_{\rm{B}}$)~\cite{RAgSe2}.
The Weiss temperature estimated from the simple Curie-Weiss fitting of the magnetic susceptibility below 300~K is $- 71.9$~K, indicating the predominance of the antiferromagnetic interactions, similar to YbCuS$_2$.
Although the magnetic specific heat exhibits a lambda-type anomaly at the magnetic transition temperature $T_{\rm{N}} \sim 1.8$~K, the magnetic entropy at $T_{\rm{N}}$ is only $30\%$ of $R \ln 2$ expected for the crystalline electric field ground-state doublet, indicating the presence of the magnetic fluctuations above $T_{\rm{N}}$~\cite{RAgSe2}.
As $T_{\rm{N}}$ is much smaller than the Weiss temperature and a peculiar temperature-magnetic field phase diagram similar to that observed in YbCuS$_2$ was obtained~\cite{YbAgSe2},
magnetic frustration due to the Yb zigzag chains is expected.

In this paper, we report the magnetic ordered state and low-energy magnetic excitations investigated from microscopic point of view with the $^{77}$Se-nuclear magnetic resonance (NMR).
Below $T_{\rm N}$, the NMR spectral intensity suddenly decreases, and the sharp and broad signals were observed.
This indicates that the internal magnetic fields at the two crystallographically distinct sites below $T_{\rm{N}}$ are different from each other.
The nuclear spin-lattice relaxation rate $1/T_1$ stays almost constant above $T_{\rm{N}}$ and suddenly decreases below $T_{\rm{N}}$, suggesting a gapped magnetic ordered state. 
$T$-linear behavior observed in YbCuS$_2$ was not detected down to 1.0~K in YbAgSe$_2$.

\section{Experimental}

Polycrystalline samples of YbAgSe$_2$ were synthesized by melting the constituent elements in evacuated quartz ampoules~\cite{RAgSe2}.
A conventional spin-echo technique was used for the $^{77}$Se-NMR measurements.
$^{77}$Se nucleus, the natural abundance of which is 7.6 \%, has nuclear spin $I = 1/2$ with nuclear gyromagnetic ratio of $^{77}\gamma/2\pi$ = 8.12 MHz/T.
A $^3$He-$^4$He dilution refrigerator was used for the NMR measurement below 1.4~K.
The $^{77}$Se-NMR spectra were obtained by the $H$-swept measurement.
The Knight shift $K$ was measured at fixed frequencies of 24.4 ($ \sim 3$~T) and 84.12~MHz ($ \sim 11$~T). 
Since it is difficult to separate the signals arising from the two sites in the spectrum measured at 24.4~MHz, the average value of Knight shift $K_{\rm av}$ was estimated from the first moment of the NMR spectrum.  
The nuclear spin-lattice relaxation rate $1/T_1$ was evaluated by fitting the relaxation curve of the nuclear magnetization after its saturation to a theoretical function for the nuclear spin $I = 1/2$, which is a single exponential function on the polycrystalline sample.
1/$T_1$ was measured at 3~T (22.4~MHz) above 1.5~K and 1~T (9.35~MHz) below 3~K.

\section{Results and Discussion}

\begin{figure*}[t!]
\centering
\includegraphics[width=17cm]{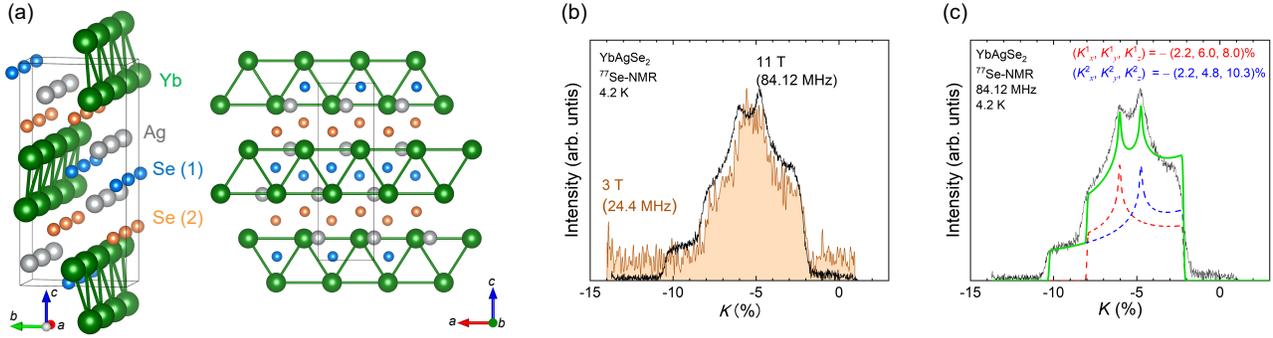} 
\caption{(Color online) (a) Crystal structure of YbAgSe$_2$ drawn by VESTA~\cite{vesta}.
(b) The $^{77}$Se-NMR spectrum obtained by $H$-swept method at 4.2 K at frequencies 24.4~MHz (brown) and 84.12~MHz (black) as a function of $^{77}$Se-NMR Knight shift.
(c) The simulation of the $^{77}$Se-NMR spectrum with two Se site with Knight shift $(K_x^1, K_y^1, K_z^1) = -(2.2, 6.0, 8.0)\%$ and $(K_x^2, K_y^2, K_z^2) = (2.2, 4.8, 10.3)\%$.}
\label{simulation}
\end{figure*}

Figure~\ref{simulation}(b) shows the $^{77}$Se-NMR spectrum obtained by $H$-swept method at the frequencies of $f = 24.4$~MHz and 84.12~MHz at 4.2~K on the powdered YbAgSe$_2$, where the horizontal axis is $K=(2\pi f/\gamma-H)/H$.
While the spectrum measured at a higher frequency has a higher resolution, NMR spectra at both frequencies are nearly identical and $H$-independent.
Since YbAgSe$_2$ has two different Se sites as shown in the Fig.~\ref{simulation}(a)~\cite{YbAgSe2_structure}, two $^{77}$Se-NMR signals are considered to overlap with each other.
The NMR spectrum was well fitted by the simulation assuming the two different Se sites with negative Knight shifts and three-axis anisotropy $(K_x^1, K_y^1, K_z^1) = -(2.2, 6.0, 8.0)$\% and $(K_x^2, K_y^2, K_z^2) = -(2.2, 4.8, 10.3)$\%, respectively. 
Here, $K_i^j$  is $i$-axis Knight shift at the Se site $j$ ($i = x, y, z$, $j = 1, 2$).
As the values of $K_i^1$ and $K_i^2$ are close to each other, the site assignment is difficult in the present NMR measurements.

\begin{figure*}[t!]
\centering
\includegraphics[width=13cm]{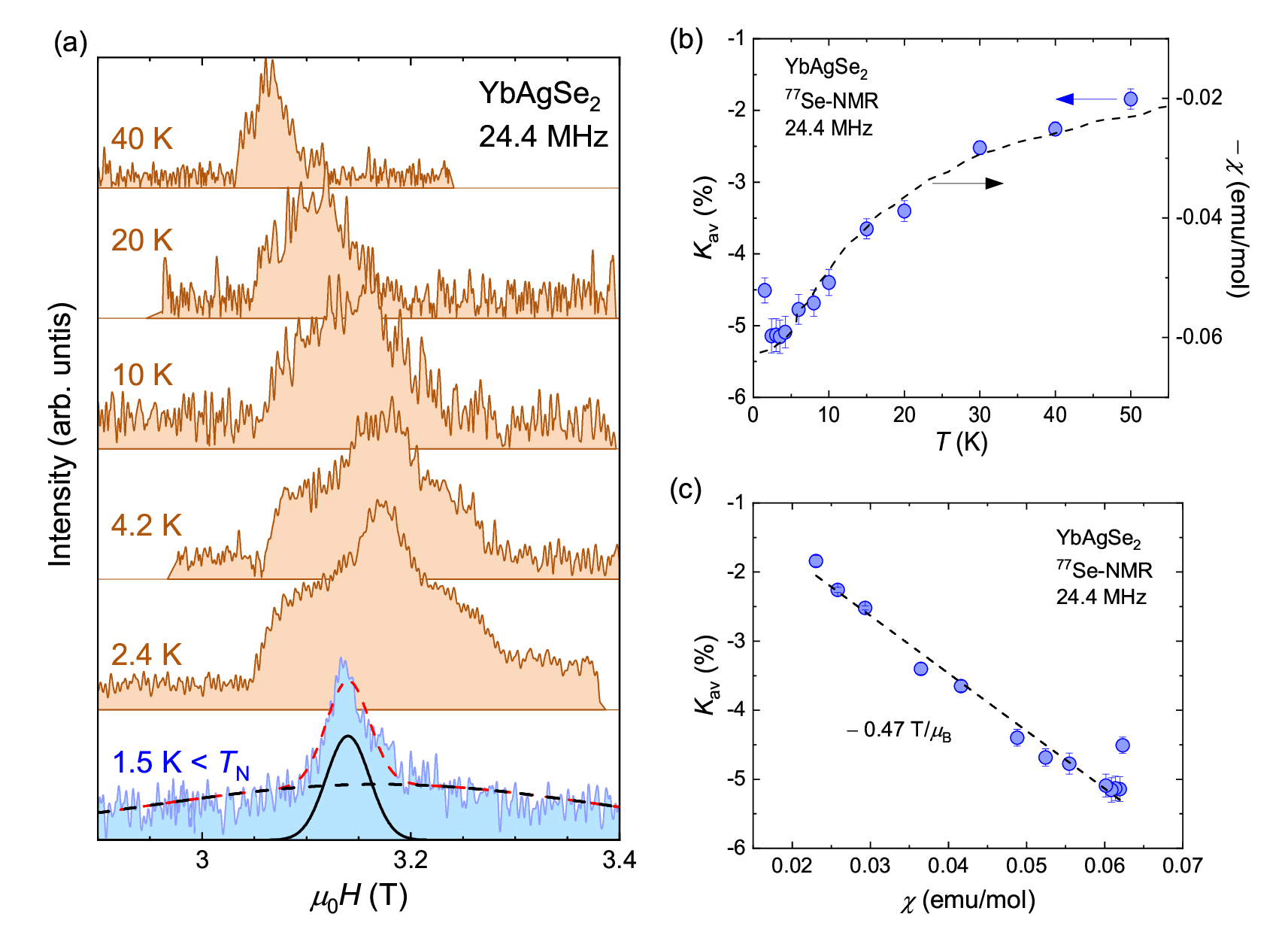} 
\caption{(Color online) (a) Temperature variations of the $^{77}$Se-NMR spectrum obtained by $H$-swept method at a frequency of 24.4~MHz for $1.5 \leq T \leq 50$~K.
(b) Temperature dependence of the average of $^{77}$Se-NMR Knight shift $K_{\rm av}$ and bulk magnetic susceptibility $\chi$.
(c) $K$-$\chi$ plot with the temperature as an implicit parameter.}
\label{Knight_shift}
\end{figure*}

As shown in Fig.~\ref{Knight_shift}(a), the NMR spectrum becomes broader and shifts toward the higher magnetic fields above $T_{\rm N}$ with decreasing temperature.
Figure~\ref{Knight_shift}(b) exhibits the temperature dependence of $K_{\rm av}$ measured at 24.4~MHz. 
Above $T_{\rm N}$, the absolute value of $K_{\rm av}$ increases with decreasing temperature, and follows the Curie-Weiss behavior. $K_{\rm av}$ is scaled with the bulk magnetic susceptibility $\chi$.
From the $K_{\rm av}$-$\chi$ plot
with $\chi$ measured at 0.1~T~\cite{RAgSe2}, the hyperfine coupling constant $A_{\rm{hf}}$ was estimated to be $- 0.47$~T/$\mu_{\rm{B}}$ as shown in Fig.~\ref{Knight_shift}(c).
Here, the hyperfine coupling constant is the proportionality constant for the Knight shift with respect to magnetic susceptibility, and thus can be obtained from the slope of the $K$-$\chi$ plot.
Such a negative hyperfine coupling constant was often observed in several Yb-based systems\cite{YbP, YbRh2Si2, YbPtGe2, YbAl3C3,  YbAlB4}.
Below $T_{\rm{N}} \sim 1.8$~K, the spectral intensity decreases.
It seems that one of the NMR spectra becomes broadened below $T_{\rm N}$, the signal intensity of which is too weak to be observed in the whole region [black dashed curve of Fig.~\ref{Knight_shift}(a)], and that the relatively sharp peak signal remains [black solid curve of Fig.~\ref{Knight_shift}(a)].
Therefore, the internal magnetic fields at the two crystallographically distinct sites below $T_{\rm{N}}$ are different from each other: one site is strongly affected by the internal magnetic fields and the other is not.
Determining which Se site is more strongly affected by the internal magnetic fields requires information about the magnetic structure.
Thus, in order to determine the magnetic structure, the neutron scattering measurements are necessary.

\begin{figure}[t!]
\includegraphics[width=8cm]{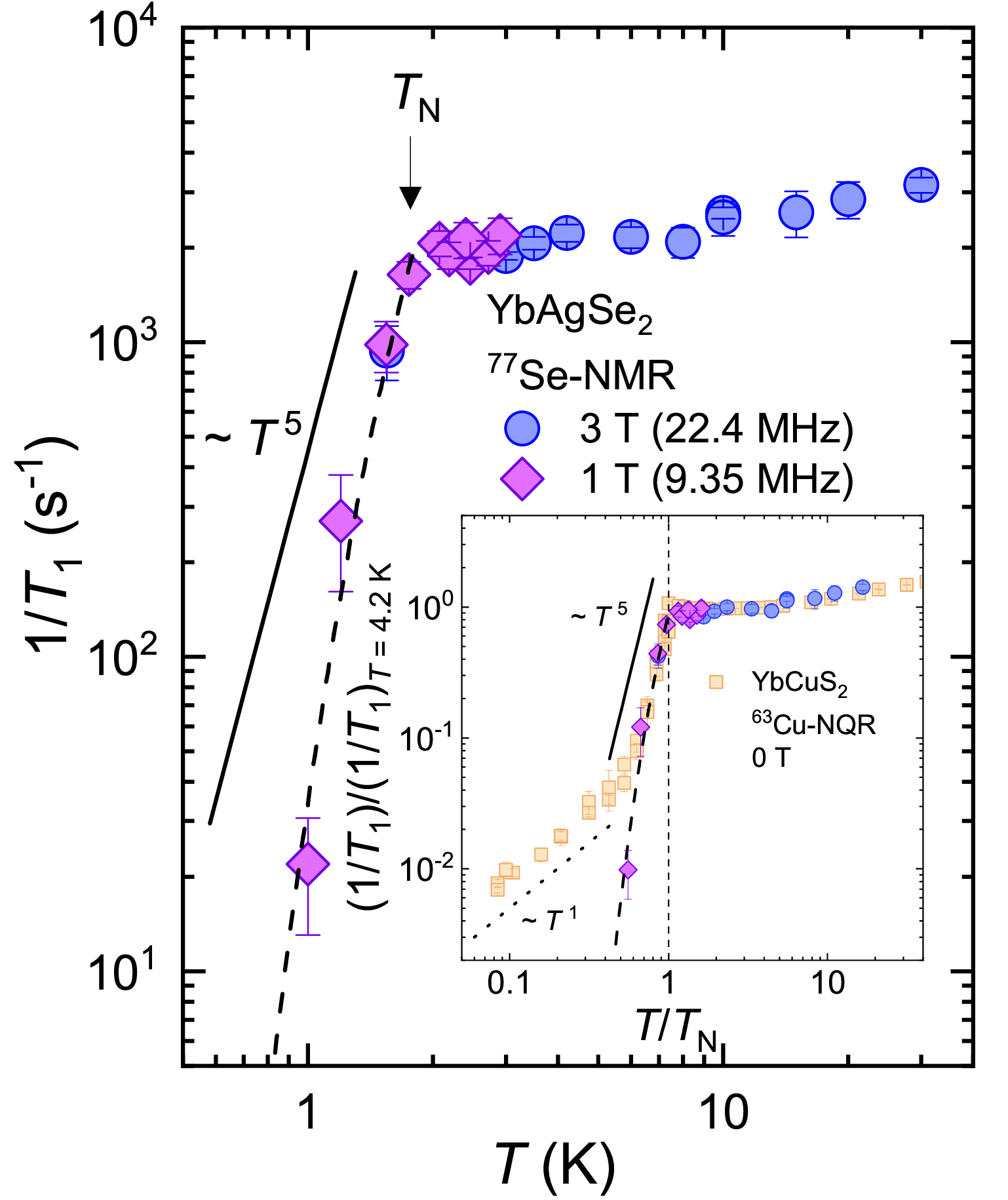} 
\caption{(Color online) Temperature dependence of the $^{77}$Se-NMR nuclear spin-lattice relaxation rates 1/$T_1$ in YbAgSe$_2$: the circles and diamonds denote 1/$T_1$ measured in 3~T (22.4~MHz) and 1~T (9.35~MHz), respectively. The inset shows $(1/T_1)/(1/T_1)_{T = 4.2~\rm{K}}$ as a function of $T/T_{\rm{N}}$: squares donates $^{63}$Cu-NQR 1/$T_1$ measured on YbCuS$_2$.~\cite{Hori2023}}
\label{1_T1}
\end{figure}

Main figure of Fig.~\ref{1_T1} shows the nuclear spin-lattice relaxation rate $1/T_1$ measured at the peak of the $^{77}$Se-NMR spectrum on YbAgSe$_2$. 
$1/T_1$ remains almost constant above $T_{\rm{N}}$, which is usually observed in localized spin systems.
Below $T_{\rm{N}}$, $1/T_1$ decreases more rapidly than $T^5$ dependence.
Tentatively, when the observed sharp decrease in $1/T_1$ down to 1.0~K is fitted to a thermally activated behavior $1/T_1 \sim \exp(-\Delta/T) $ with a magnon gap $\Delta$, it is obtained as $\Delta \sim 9$~K, which is larger than the transition temperature $T_{\rm N}$.

Here, we discuss the hyperfine coupling constant and low-energy excitations of YbAgSe$_2$ in comparison with those of YbCuS$_2$.
The absolute value of $A_{\rm{hf}}=-0.47$~T/$\mu_{\rm{B}}$ in YbAgSe$_2$, determined from the slope of the $K$-$\chi$ plot, is three times larger than $0.14$~T/$\mu_{\rm{B}}$ in YbCuS$_2$~\cite{Hori_NMR}.
In general, hyperfine coupling constant in insulators or semiconductors originates from the classical dipolar interaction.
However, $A_{\rm{hf}}$ evaluated in YbAgSe$_2$ is much larger than the hyperfine coupling constant estimated from the classical dipolar interaction $A_{\rm d}$, which is given by
\begin{flalign}
\begin{aligned}
A_{\rm d}&=\sum_{i} 9.2741 \times 10^{-1} \\
&\times \frac{1}{r_i^{5}}\left(\begin{array}{ccc}
3 x_{i}^{2}-r_i^{2} & 3 x_i y_i & 3 x_i z_i \\
3 y_i x_i & 3 y_i^{2}-r_i^{2} & 3 y_i z_i \\
3 z_i x_i & 3 z_i y_i & 3 z_i^{2}-r_i^{2}
\end{array}\right) \;\; \left({\rm T}/ \mu_{B}\right),
\end{aligned}
\end{flalign}
where $r_i$~($\AA$) is a distance between Se and Yb-$i$ sites.
From the eq. (1),
\begin{flalign}
{A}_{\rm d}^{\rm{Se(1)}}= \left(\begin{array}{ccc}
0.0046 & -0.0027 &  0.0051 \\
-0.0027& 0.0018& 0.0082 \\
0.0051 & 0.0082 & -0.0064
\end{array}\right) \;\; \left({\rm T} / \mu_{B}\right)
\end{flalign}
for the Se(1) site, and
\begin{flalign}
{A}_{\rm d}^{\rm{Se(2)}}= \left(\begin{array}{ccc}
0.0078 & 0.0029 & -0.0030 \\
0.0029& -0.0097& -0.1023 \\
-0.0030 & -0.1023 & 0.0018
\end{array}\right) \;\; \left({\rm T} / \mu_{B}\right)
\end{flalign}
for the Se(2) site are obtained.
In addition, as shown above, all components of Knight shift are negative, which is not explained with the dipole interaction solely. 
Therefore, the hyperfine coupling constant is not mainly determined by the dipole interaction.
One possible origin is the core polarization effect induced by $p$ electrons of the Se atoms, which gives a negative Knight shift.
Note that the core polarization effect is typically associated with $d$ electrons\cite{Pt-NMR, Pt-Rh, Sr2RuO4, NaCo2O4}, and there are a few examples of the core polarization by $p$ electrons\cite{EuNi2P2_Magishi, EuNi2P2_Higa}.


Next, we discuss the difference of $1/T_1$ between Yb-zigzag chain systems.
Note that the value of $1/T_1 \sim 2000$~s$^{-1}$ at around 4.2~K for YbAgSe$_2$ is more than one order of magnitude larger than that for YbCuS$_2$ ($1/T_1 \sim 150$~s$^{-1}$).
The difference seems to be consistent with the one in the hyperfine coupling constant.
In order to compare the $1/T_1$ behavior of YbCuS$_2$ and YbAgSe$_2$, $1/T_1$ was normalized at a value of 4.2~K, and the temperature was normalized with $T_{\rm N}$ as shown in the inset of Fig.~\ref{1_T1}. 
Above $T_{\rm N}$, the behavior of $1/T_1$ in YbAgSe$_2$ is almost the same as that of YbCuS$_2$ and the critical slowing down behavior was not observed near $T_{\rm{N}}$ on both compounds. 
In previous study~\cite{Hori2023}, we reported that the absence of the critical slowing down behavior is likely to be related to the character of the first-order phase transition in YbCuS$_2$.
However, this contradicts with the experimental result that the specific heat of YbAgSe$_2$ exhibits a lambda-type anomaly at $T_{\rm{N}}$, indicating a second-order transition~\cite{RAgSe2}.
Therefore, such a temperature dependence of $1/T_1$ might be a unique behavior on the Yb zigzag chain systems, but the detailed origin remains unclear and further investigations are required.

Below $T_{\rm N}$, YbAgSe$_2$ shows a gapped behavior in $1/T_1$ and $T$-linear behavior observed in YbCuS$_2$ was not detected down to 1.0~K. 
There are three possibilities to explain the absence of the $T$-linear behavior in YbAgSe$_2$. 
One possibility is that the gapless behavior is unique to YbCuS$_2$, although both compounds possess the similar Yb zigzag chains.
First order character of the magnetic transition might be related to the gapless excitation.
Another possibility is that the gapless behavior might be observed at lower temperatures than 1.0~K in YbAgSe$_2$. 
However, the intensity of the $^{77}$Se-NMR spectrum in YbAgSe$_2$ becomes so weak, and $T_1$ becomes so long below $T_{\rm N}$, which give limitations on the temperature range for $^{77}$Se-NMR measurements. 
Currently, NMR measurements can only be performed down to 1.0~K in this compound.
The other possibility is that the gapless behavior in YbCuS$_2$ might be immediately suppressed against magnetic field, because the previous measurements in YbCuS$_2$ was performed at zero magnetic field.
Since Se nuclei have $I = 1/2$, NQR measurements at zero magnetic field are impossible on YbAgSe$_2$.
Thus, to verify this possibility, $^{63}$Cu-NMR $1/T_1$ measurements on YbCuS$_2$ under magnetic fields or zero-field muon spin resonance in YbAgSe$_2$ is needed, which is now in progress.

\section{Conclusion} 

In this study, we performed the $^{77}$Se--NMR studies on Yb zigzag chain compound YbAgSe$_2$ to compare to YbCuS$_2$.
The $^{77}$Se-NMR spectrum was fitted by simulating two different Se sites with negative Knight shifts and three-axis anisotropy.
The Knight shift is scaled to the bulk magnetic susceptibility above $T_{\rm N}$.
Below $T_{\rm N}$, the spectral intensity decreases, and the relatively sharp signal and the extremely-broad signal with weak intensity coexist with each other.
This observation indicates a difference in the internal fields at the two crystallographically distinct sites below $T_{\rm{N}}$: one site is affected by the internal magnetic fields and the other is not.
$1/T_1$ stays almost constant above $T_{\rm{N}}$ and follows $1/T_1 \sim \exp (-\Delta/T )$ below $T_{\rm{N}}$, suggesting a magnetic ordered state with magnon gap, but the $T$-linear behavior of $1/T_1$, which suggests the gapless excitation, was not observed down to 1.0 K. 
This is in contrast to what was observed in YbCuS$_2$, where the $T$-linear behavior was observed below 0.5~K.
It is possible that $T$-linear behavior is unique to YbCuS$_2$, or might be immediately suppressed under the magnetic fields.

\begin{acknowledgments}
The authors would like to thank S. Yonezawa, H. Saito, H. Nakai, and C. Hotta for valuable discussions. 
This work was supported by Grants-in-Aid for Scientific Research (KAKENHI Grant No. JP20KK0061, No. JP20H00130, No. JP21K18600, No. JP22H04933, No. JP22H01168, No. JP23H01124, No. JP23K22439, No. JP23K25821, No. JP23H04866, No. JP23H04870, No. JP23KJ1247 and No. JP24K00574) from the Japan Society for the Promotion of Science, by JST SPRING (Grant No. JPMJSP2110) from Japan Science and Technology Agency, by research support funding from the Kyoto University Foundation, by ISHIZUE 2024 of Kyoto University Research Development Program, and by Murata Science and Education Foundation.
In addition, liquid helium is supplied by the Low Temperature and Materials Sciences Division, Agency for Health, Safety and Environment, Kyoto University.
\end{acknowledgments}

\end{document}